\documentclass[conference]{IEEEtran}
\IEEEoverridecommandlockouts

\usepackage{cite}
\usepackage{amsmath,amssymb,amsfonts}
\usepackage{algorithmic}
\usepackage{graphicx}
\usepackage{textcomp}
\usepackage{xcolor}
\usepackage{mdframed}
\def\BibTeX{{\rm B\kern-.05em{\sc i\kern-.025em b}\kern-.08em
    T\kern-.1667em\lower.7ex\hbox{E}\kern-.125emX}}
\usepackage{hyperref}
\usepackage{cleveref}
\usepackage{enumitem}
\usepackage{tabularx}
\usepackage{booktabs}
\usepackage{array}

\usepackage{flushend}

\newlist{tabitemize}{itemize}{1}
\setlist[tabitemize]{label=\textbullet, nosep, leftmargin=*, after=\vspace{-.2\baselineskip}, before=\vspace{-0.5\baselineskip}}

\begin{document}

\title{Information is all you need:\\Requirements Engineering Quality Reframed}

\author{
    \IEEEauthorblockN{Henning Femmer}
    \IEEEauthorblockA{
        \textit{South Westphalia University of Applied Sciences}\\
        Hagen, Germany \\
        femmer.henning@fh-swf.de}
    \and
    \IEEEauthorblockN{Julian Frattini}
    \IEEEauthorblockA{
        \textit{Chalmers and University of Gothenburg}\\
        Gothenburg, Sweden \\
        julian.frattini@chalmers.se}
}

\maketitle

\begin{abstract}
     Why do low-quality requirements specifications sometimes produce functional software while some projects with high-quality specifications fail?
     A common answer to this anomaly is that \emph{it depends on the context}, which is neither expressive nor actionable.
     To move beyond this vague appeal to context, this vision proposes a novel holistic theory of requirements engineering (RE) quality. 
     This theory models how information particles, i.e., discrete pieces of domain knowledge, are transferred between roles and artifacts.
     Since the RE process is ultimately an information transfer, holistic RE quality depends on the properties of information flow, i.e., how effectively and efficiently information is transferred from sources (like stakeholders) to targets (like developers and testers),
     uniting both artifact- and process-based perspectives on RE quality.
     In an exemplary simulation of the theory we illustrate why a high-quality specification gets bypassed in an agile context, thereby demonstrating that a simulation can provide actionable insights into calibrating the RE process to optimize the information flow.
     Beyond organizational applications, we envision that the theory can serve as a coherent theoretical framework for understanding the success or failure of RE processes and artifacts.
\end{abstract}


\section{Introduction}

In the software engineering (SE) process, various studies stress the relevance of requirements engineering (RE) for project success~\cite{fernandez2017naming}. 
Against many claims, challenges in RE persist regardless of the adopted process model (e.g., waterfall or agile)~\cite{fernandez2017naming}.
These challenges manifest when root causes such as incomplete or hidden requirements~\cite{fernandez2017naming} affect subsequent activities that rely on these requirements as input~\cite{femmer2015activities}, e.g., implementing or testing features.
As a consequence, various RE quality models have postulated how (not) to write requirements specifications~\cite{iso2018systems,cpre2024,femmer2017rapid,phalp2007assessing} to avoid such impacts.

However, the adequacy of these quality models has been drawn into question~\cite{frattini2023requirements}.
Despite plausible arguments, several factors of requirements quality experience contradictory evidence, where some empirical studies claim that factors such as \emph{passive voice}~\cite{femmer2014impact} or \emph{consistently numbered steps in use cases}~\cite{phalp2007assessing} matter, while others disagree~\cite{krisch2015myth,frattini2025adopting}.
The RE community seems content with the consensus to explain these differences with the ominous influence of context~\cite{frattini2023requirements}.
Deflecting the explanation to context is not helpful, as it does not provide reliable decision support for addressing these differences.

These differences resisting explanation by existing quality theories constitute an anomaly requiring a paradigm shift~\cite{Kuhn2012-ox}.
To this end, we propose an alternative paradigm to the current artifact-centric perspective to RE quality.
Consistent with information theory~\cite{durugbo2013modelling}, we frame RE as a means of information transfer:
Information in the form of requirements needs to be transferred from its sources (e.g., stakeholders, competitor analyses, reuse databases) to its targets just beyond the border of RE (e.g., designers, developers, and testers).
RE quality can then be understood as how effectively and efficiently information flows, regardless of its form (i.e., written or verbal), offering a holistic perspective on quality that goes beyond artifact- and process-centric theories.

Our vision for such an information-centric theory of RE quality outlines an avenue of research capable of resolving the aforementioned anomalies.
The resulting theory can both explain them, and be operationalized to mitigate them.
In the scope of this vision paper, we make the following contributions:

\begin{itemize}
    \item A novel paradigm framing RE quality from a holistic perspective, focusing on the flow of information particles.
    \item A simulation of the application of this paradigm to demonstrate its envisioned use case (available online~\cite{renext26replication}). 
\end{itemize}

\section{Theoretical Foundations}
\label{sec:foundation}

In this section, we introduce the relevant background underpinning this vision. 
Our presentation follows the epistemological framework proposed by Kuhn~\cite{Kuhn2012-ox}, according to which research is governed by predominant paradigms that persist until accumulated anomalies can no longer be reconciled with them, triggering a shift toward a more powerful theory.

\Cref{sec:foundation:quality} elaborates the current paradigm of quality in RE, and \Cref{sec:foundation:anomaly} provides evidence for current anomalies resisting explanation.
Finally, we introduce related work on information flow in \Cref{sec:foundation:information}.

\subsection{Requirements Quality: Current Paradigms}
\label{sec:foundation:quality}

Today, we understand RE either from a perspective of (physical) artifacts or from a perspective of processes and techniques. 
Most approaches have their own definition of quality without a higher theoretical foundation.

Most fundamentally, early works~\cite{pohl1994three,lindland1994understanding,krogstie1995towards,krogstie1998integrating} iteratively defined RE quality as relationships between different concepts. 
Quality is ultimately understood as a relationship between the artifact and the domain (\emph{semantic quality}), the RE language (\emph{syntactic quality}) and the audience interpretation (\emph{pragmatic quality}). 
Further qualities include the consistency of the audience interpretation (\emph{social quality}) and various types of appropriateness, e.g., the choice of language for the intended audience~\cite{krogstie1998integrating}.

Common quality models used in standards---such as ISO/IEC/IEEE 29148:2018~\cite{iso2018systems} or the IREB CPRE FL~\cite{cpre2024}--- focus on sets of generalized quality factors (such as completeness, unambiguity, etc.), identified and specified by experts in the field. 
The way these factors are postulated, they usually do not provide falsifiable reasoning about the consequences of not adhering to the standard, and accordingly, are presented without empirical underpinning.
To amend this, follow-up research postulated artifact-centric theories that determine requirements quality based on artifacts' impact on subsequent activities~\cite{femmer2018requirements,femmer2015activities,frattini2023requirements}, leading to falsifiable quality factors:
An artifact without negative impact on subsequent activities cannot be considered defective.

More specific definitions for quality exist for specific artifact types such as user stories~\cite{jeffries2001essential,cohn2004user,lucassen2016improving}, use cases~\cite{phalp2007assessing}, vision videos~\cite{KARRAS2020110479}, or UML models~\cite{lange2006improving}. 
Similarly, authors came up with quality criteria for different techniques, e.g., for interviewing~\cite{bano2018learning}, or conducting workshops~\cite{gottesdiener2002requirements}.

\subsection{Persisting Anomalies}
\label{sec:foundation:anomaly}

The artifact-centric quality theories presented in \Cref{sec:foundation:quality} are based on plausible arguments, yet \emph{anomalies}---observable phenomena that resist explanation by these theories~\cite{Kuhn2012-ox}---still persist.
Particularly, when empirical studies contribute evidence to the theories, contradictions arise.

For example, several prior studies have suggested that textual requirements using passive voice are defective~\cite{parra2015methodology} because it ``can lead to ambiguous interpretations''~\cite{ferrari2018detecting}.
Femmer et al.~\cite{femmer2014impact} observed in a controlled experiment that passive voice in textual requirements affects the reader's ability to relate domain concepts to each other supporting this claim.
However, a later replication did not confirm this effect~\cite{frattini2025applying}.
Even more contradictory, Krisch and Houdek arrived at the opposite conclusion during a case study, i.e., that practitioners consider most occurrences of passive voice harmless because ``[i]nformation can be drawn through context''~\cite{krisch2015myth}.

Similarly, catalogs of alleged requirements quality defects see varying support from empirical studies.
For example, the 7 C's by Phalp et al.~\cite{phalp2007assessing} contain  writing rules with a strong foundation in text comprehension theory, but evidence from an empirical case study shows only partial support for them~\cite{frattini2025adopting}.

A \emph{context factor} is often cited as a kind of \textit{deus ex machina} to explain why  quality sometimes does (not) matter~\cite{krisch2015myth,frattini2023requirements}:
An occurrence of a specific quality factor may be harmful in one context but irrelevant in another.
While plausible, this coarse explanation is unhelpful to properly understand or act on the alleged defect, though, as it is limited to claiming \emph{that} context affects quality~\cite{mund2015does,frattini2023requirements}, but not \emph{how}.
This remains a limitation of existing RE quality theories.

\subsection{Related Work on Information Flow in Companies}
\label{sec:foundation:information}

Our theory---introduced in \Cref{sec:paradigm}---is based on the idea that requirements are information flowing through a socio-economic system. 
Although we could not find a fitting theory that can be reused in the RE context, researchers in the fields of information systems and economics have studied concepts of information and information flow from different angles.

In times of information-based organizations~\cite{drucker1988coming}, the organizations' task is to balance information supply and information demand, i.e., a dynamic system that requires continuous adjustment~\cite{krcmar2015informationsmanagement}. 
In this context, information is described as a specific good: immaterial, useful, valuable, mutable, reproducible, and transportable at the speed of light~\cite{krcmar2015informationsmanagement}.

Durugbo et al. have summarized various approaches for modeling the flow of information in organizations~\cite{durugbo2013modelling}.
Diagrammatic modeling allows organizations to better understand their own information flow on a high level~\cite{juric1999engineering} and falls into three categories:
pictorial representations~\cite{burstein2004framework}, graph representations~\cite{feinstein1988information,braha2007statistical}, and matrix representations~\cite{al2008modelling}.
Central to diagrammatic information flow modeling are often the nodes (i.e., between whom the information flows) rather than the actual information itself.

The alternative mathematical modeling approaches serve two main purposes: flow analysis and organizational analysis~\cite{durugbo2013modelling}.
Krovi et al. propose a ``parameter-based guiding framework of information flow to manage organizational processes''~\cite{krovi2003information} drawing inspiration from fluid flow with concepts like velocity and viscosity, but remain on a descriptive level.
Lin and Cheng~\cite{lin2007structural} formalize organizations as interrelated parts whose behavior is defined by mathematical functions, but focuses rather on the overall relationship flow within an organization.
Ben Arieh and Pollatscheck~\cite{ben2002analysis} mathematically describe information overload in hierarchical organizations, modeling organizations as trees mainly with interactions along the hierarchy.

\subsection{Related Work on Information Flow in RE}
Winkler~\cite{sawyer_information_2007} analyzes the information flow (only) between documents, acknowledging that people do not always keep the tangible documents consistent with the intangible knowledge.

Schneider et al.~\cite{schneider_beyond_2008} model flow of information in RE using various techniques, such as UML diagrams or data flow diagrams. The paper distinguishes between solid and fluid representations, which closely resembles our tangible and intangible information storage concepts. 
They also introduce the FLOW notation to explicitly model both types. While their work focuses more on flow and visualization, our work has a focus on the concept of the information itself, its simulation, and what this theoretic lens implies for RE quality.

Despite their relatedness, none of the existing approaches matches our use case.
Diagrammatic modeling focuses on the nodes between which information flows whereas our perspective focuses on the flowing information itself.
Mathematical modeling approaches are either non-formal~\cite{krovi2003information} or focus on specific aspects~\cite{lin2007structural,ben2002analysis}

\section{An Information-Centric Paradigm of RE Quality}
\label{sec:paradigm}

The anomalies presented in \Cref{sec:foundation:anomaly} resist explanation by the current paradigms of quality in RE presented in \Cref{sec:foundation:quality}.
To resolve this crisis~\cite{Kuhn2012-ox}, we propose a novel paradigm of RE quality based on information flow introduced in \Cref{sec:foundation:information}.
First, we elaborate a deliberately simplified yet plausible RE scenario in \Cref{sec:paradigm:anomaly} to illustrate the concepts then introduced in \Cref{sec:paradigm:concepts}.
With these concepts, we resolve the described anomaly in \Cref{sec:paradigm:mapping}.

\begin{figure*}
    \centering
    \includegraphics[width=\linewidth]{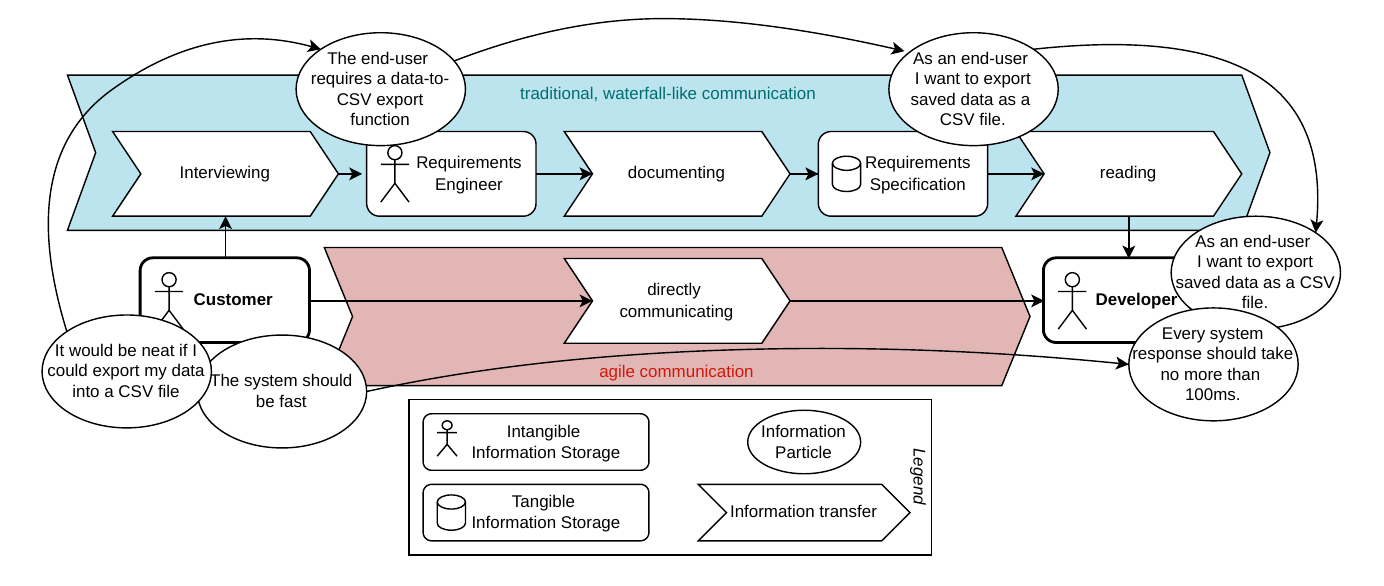}\vspace{-1em}
    \caption{Synthetic scenario showing two alternative paths for information flow}
    \label{fig:anomaly}
\end{figure*}

\subsection{The Anomaly}
\label{sec:paradigm:anomaly}

Assume the following plausible, albeit simplified scenario while observing software developers in a small project team. 
A skilled requirements engineer has created a high-quality requirements specification that stakeholders have approved. 
It is comprehensive, complete, and consistent, i.e., it meets all criteria commonly demanded of a requirements specification~\cite{iso2018systems,cpre2024,phalp2007assessing}.
Despite this, the developers consistently bypass it and talk directly to the customer instead. 

This behavior is inexplicable by existing requirements quality theories:
Adhering to requirements writing guidelines~\cite{iso2018systems,cpre2024,phalp2007assessing} should maximize the usability of the requirements specification according to artifact-centric quality theories~\cite{femmer2018requirements}.
Hence, these theories do not rationally explain the divergent behavior.
The most common answer for dodging this anomaly is deferring to dependency on the context~\cite{mund2015does}.
But even reference to existing catalogs of context factors~\cite{petersen2009context} do not sufficiently resolve the anomaly.

\subsection{Core Concepts of an Information-Centric RE Paradigm}
\label{sec:paradigm:concepts}

The reason for this anomaly is rooted in the structure of the existing theories itself.
Artifact-centric RE quality theories consider \emph{requirements artifacts} (e.g., requirements specifications, user stories, or backlogs) as first-class citizens.
But requirements artifacts are just one means-to-an-end, i.e., one form of how requirements can be stored and transferred, just as RE itself is a means-to-an-end~\cite{cpre2024}.
This end, i.e., the ultimate purpose of RE, is gathering requirements from stakeholders and other sources, and providing them to the SE process where a solution to these requirements is developed.
In the following, we frame this purpose of RE from an information-centric perspective to arrive at a holistic paradigm of RE quality.

\paragraph{Information particles and flow}

Requirements are information, particularly information about the problem-space~\cite{nuseibeh2001weaving,fernandez2012field}.
A single requirement, i.e., a single, atomic piece of information, can be considered an \textbf{information particle}~\cite{krovi2003information}.
For example, a need expressed by a customer about a certain feature (e.g., the feature request to export data in CSV format, see \Cref{fig:anomaly}) is one such information particle.
The semantic information, syntactically and lexically refined during the RE process, later informs development about the features to implement. 

Information particles are located in \textbf{information storages}.
Contrary to existing theories~\cite{femmer2015activities,frattini2023requirements}, these storages can both be tangible (e.g., requirements specifications, user story backlogs, basically any RE artifact) or intangible (e.g., a stakeholder's prior domain knowledge or a developer's memory), as shown in \Cref{fig:anomaly}.
The capacity and retention of tangible and intangible storages may differ, but they both serve the same purpose from an information-centric perspective, i.e., storing information particles.

Information particles can move between information storages during \textbf{information transfers}.
Any activity that obtains information particles from one storage and moves it to another qualifies as an information transfer.
For example, while \textit{interviewing}, information particles will transfer from a customer to the requirements engineer (top left in \Cref{fig:anomaly}).
Afterwards, while \textit{documenting}, this requirements engineer transfers the obtained information particles to a requirements specification.
How much and how quickly information can be transferred from one storage to another depends on factors like task complexity and skill~\cite{arisholm2007evaluating}.

Overall, the RE phase is a sequence of information transfers that move information particles from source information storages (e.g., stakeholders, competitor analyses, or field studies) to target information storages (e.g., software architects, developers, or testers), potentially via multiple different paths.
Target information storages are those located just outside the boundary of RE, i.e., where problem-space information (i.e., requirements) are used to produce solution-space information (e.g., architecture, source code).
For example, the \textit{developer} in \Cref{fig:anomaly} is an information storage outside of boundary of RE, as they use the problem-space information to implement the required features.
Overall, the RE phase is---from an information-centric perspective---simply an \textbf{information flow}~\cite{krovi2003information}.

\paragraph{RE quality in the information-centric paradigm}

Given this framing, the \emph{quality of RE} becomes the \textbf{effectiveness and efficiency} of the information flow, i.e., how much information arrives at the boundary of RE and how much resources (e.g., time) it costs.
This definition subsumes previous requirements quality theories.
In particular, the previous artifact-centric requirements quality theories are completely encompassed.
Artifact-centric theories explain quality via the impact of requirements artifacts on subsequent activities~\cite{femmer2015activities}, which maps to the effectiveness and efficiency of information transfers originating from tangible information storages.
However, our proposed paradigm embeds this relationship in a larger-scale, holistic perspective where these effects co-exists with information transfers originating from intangible information storages.

\subsection{Resolving the Anomaly}
\label{sec:paradigm:mapping}

The paradigm elaborated in \Cref{sec:paradigm:concepts} can resolve the anomaly from \Cref{sec:paradigm:anomaly}.
In this scenario, the developer---i.e., an information storage just beyond the border of RE---requires information for the solution-design and has two ways of obtaining that information. 
They can either have the information transferred by \textit{reading} the (high-quality) requirements specification, a tangible information storage (via the top, blue route in \Cref{fig:anomaly}), or they can \textit{directly communicate} with a stakeholder, an intangible information storage (via the bottom, red route in \Cref{fig:anomaly}).
From the perspective of the developer, whichever way is more effective (i.e., yields more information) and efficient (i.e., requires less resources like time) is preferable.
So, regardless of the organization's official process model, if the developer can obtain more information more quickly by directly communicating with the customer, the overall RE quality is higher if the developer bypasses the requirements specification.
Alternatively, the customer may be more difficult to access or the developer may struggle to transfer sufficient information from the customer (e.g., because they are not trained in requirements elicitation).
In this case, direct communication yields less information particles and/or requires more effort for reaching an equivalent yield. 
Consequently, the theory will predict that the developers most likely will not bypass the requirements specification.
Overall, information follows the path of least resistance, which is how the information-centric paradigm resolves the anomaly.

However, the resolution is currently only narrative, i.e., based on plausible explanations. 
This does not suffice for evidence-based decisions, e.g., at what point of resistance the information will flow through a different path.
To enable such decisions, the paradigm requires formalization. 

\section{Formalization}
\label{sec:formalization}

Organizations could apply the paradigm of framing requirements as information particles and modeling their RE process as an information flow.
This would allow them to (1) explain previously counter-intuitive phenomena like the one presented in \Cref{sec:paradigm:anomaly}, and (2) predict how the information flow would change when calibrating the RE process, e.g., by allocating more time or higher-skilled engineers to a certain transfer.
Jointly, the application of the paradigm would serve capturing the concept of RE quality by relating RE process properties to the output of information.

Despite the existing literature on information flow modeling, no single approach directly maps to the RE context and fulfills the aforementioned goals.
We do not aim to elaborate a conclusive and definitive approach in the scope of this study, but present potential forms in \Cref{sec:formalization:approaches} and illustrate the application of one in \Cref{sec:formalization:syntax} to illustrate its usefulness.

\subsection{Approaches for Formalizing RE Information Flows}
\label{sec:formalization:approaches}

\Cref{tab:modeling} contrasts two potential forms according to their notation, advantages, and disadvantages.
The subsequent paragraphs briefly discuss each of them.

\begin{table*}[ht]
    \centering
    \small
    \caption{Comparison of Types of Formalizations of RE Information Flows}
    \label{tab:modeling}
    \begin{tabularx}{\textwidth}{p{2.5cm}|XXXX}
        \toprule
        \textbf{Type} & \textbf{Advantages} & \textbf{Disadvantages} & \textbf{Use Cases} \\
        \midrule
        
        Type I: discrete, concrete particles & 
        \begin{tabitemize}
            \item High granularity and specificity
            \item Directly empirically grounded
        \end{tabitemize} &
        \begin{tabitemize}
            \item Only retrospective
            \item Low generalizability
        \end{tabitemize} & 
        \begin{tabitemize}
            \item Retrospective analysis (e.g., identifying bottlenecks)
            \item Modeling at semantic level (e.g. introduction of inconsistencies)
        \end{tabitemize} \\

        Type II: continuous distributions &
        \begin{tabitemize}
            \item High generalizability
        \end{tabitemize} & 
        \begin{tabitemize}
            \item Requires extensive empirical evidence to construct
            \item Implies several mathematical assumptions
        \end{tabitemize} &
        \begin{tabitemize}
            \item Identification of optimal flows
            \item Prospective analysis (e.g., simulating change)
        \end{tabitemize} \\
        \bottomrule
    \end{tabularx}
\end{table*}

\paragraph{Information particles as discrete, concrete items (type~I)}

Most concretely, a formalization could describe the flow of concrete information particles directly just as exemplified in \Cref{sec:paradigm} and \Cref{fig:anomaly}.
Information particles would represent individual requirements, both functional (e.g., ``As an end-user I want to export saved data as a CSV file.'') and non-functional (e.g., ``Every system response should take no more than 100ms.'').
The formalism would model the flow of these particles from a set of information storages (e.g., customers, requirements reuse databases, or competitor analyses) through intermediate storages (e.g., requirements engineers, requirements specifications) until reaching the border of the RE process (e.g., a developer or a solution specification). 
First experiments evaluating whether information flows could be simulated with agentic AI can be found in~\cite{femmer2026sesl}.
However, for realistic projects such an information flow could currently only be determined in hindsight, i.e., once the RE phase has completed and the information particles are known.
Retrospective case studies particular to one project would support a post-mortem analysis of identifying bottlenecks and understanding properties of the specific information flow.

\paragraph{Information flow as a continuous distribution (type~II)}

To allow not only retrospective but also prospective analysis (e.g., for decision making), discrete, concrete information particles need be abstracted to a continuous quantity.
This continuous quantity would represent the \emph{level of information}, reaching from 0 (i.e., no information at all) to 1 (i.e., all relevant information), and be described in continuous distributions.
For example, abstracting from multiple cases of type~I formalizations, organizations can estimate \textit{what fraction} of the total information certain stakeholders usually contain rather than which exact information particles (which differ from project to project).
Information transfers are then defined as functions describing how much information---on average---moves from one storage to another depending on certain attributes like complexity or skill.
For example,---again abstracting from multiple type~I cases---organizations can estimate the complexity and resource-consumption of recurring transfers like \textit{interviewing} and \textit{documenting} as well as their input-output relation of information.
As the more abstract formalization type, it would be difficult to model phenomena specific to semantic properties (e.g., requirements overlap or conflict). 
On the other hand, however, this approach offers the greatest transferability of insights to other contexts.
Abstracting one type~II formalization from multiple type~I formalizations produces a generic description of the RE phase as an information flow usable for simulation and decision-making.

\subsection{Exemplary Formalization}
\label{sec:formalization:syntax}

To make the potential use case of applying the paradigm more tangible, we specify a formalization of the aforementioned type~II (information as a continuous quantity) and simulate its use.
We do not claim that this exemplary formalization is optimal for realizing type~II formalizations, but only use it for demonstration purposes.
As an overall scenario, we formalize the phenomenon described in \Cref{sec:paradigm:anomaly}, i.e., the counter-intuitive situation that a well-specified requirements artifact will be used in one context but bypassed in another.
The full details, source code, and figures can be found in our replication package~\cite{renext26replication}.

We model the information flow with two elements:

\begin{enumerate}
    \item \textbf{Information storage}, defined by the level of information $i \in [0, 1]$ that it contains
    \item \textbf{Information transfer}, defined by the level of input information $i \in [0, 1]$, the complexity level of the transfer activity $c \in \mathbb{R}^+$, the skill of the person executing the transfer $s \in \mathbb{R}^+$, the maximum time needed to completely exhaustively complete the transfer $t_{max} \in \mathbb{R}^+$, and the actual time spent executing the transfer $t \in \mathbb{R}^+$
\end{enumerate}

We  realize an information transfer $f$ as the cumulative probability distribution function of the beta distribution ($\text{CDF}_\beta$):
$$f(i, c, s, t_{max}, t) = \text{CDF}_\beta \left( \alpha=s, \beta = c, x = \frac{t}{t_{max}} \right) \cdot i$$
The $\text{CDF}_\beta$ has several properties that make it eligible to represent one information transfer.
Firstly, it produces an output between 0 and 1, which fits the quantity of \emph{level of information}.
Secondly, its shape is defined by two parameters $\alpha$ and $\beta$.
Higher values of $\alpha$ move the probability mass towards the left, i.e., increase the information output.
We define $\alpha$ with a factor representing a positive influence:
The \emph{skill} of a person is generally considered to improve the effectiveness of a task~\cite{arisholm2007evaluating}.
For example, a highly skilled requirements engineer will be able to extract more information while interviewing than an engineer with less skill.
Vice versa, higher values of $\beta$ move the probability mass towards the left, i.e., decrease the information output.
Similarly, we define $\beta$ with a factor representing a negative influence, namely the \emph{complexity} of a task.
The more complex a task, the more difficult it is to transfer information through it~\cite{arisholm2007evaluating}.
Skill and task complexity are arbitrarily chosen, but plausible and recurring context factors moderating SE phenomena~\cite{arisholm2007evaluating}.
The relationship between the invested time $t$ and the maximum time $t_{max}$ determines the investment of resources into a task.
Spending the maximum amount of time on a task results in the output of all the input information.
This is achieved by the factor $\cdot i$, which enforces that the maximum level of information produced by a task is the level of information provided as an input.
In other words: an information transfer can only lose existing, but not produce new information.

\begin{figure}
    \centering
    \includegraphics[width=\linewidth]{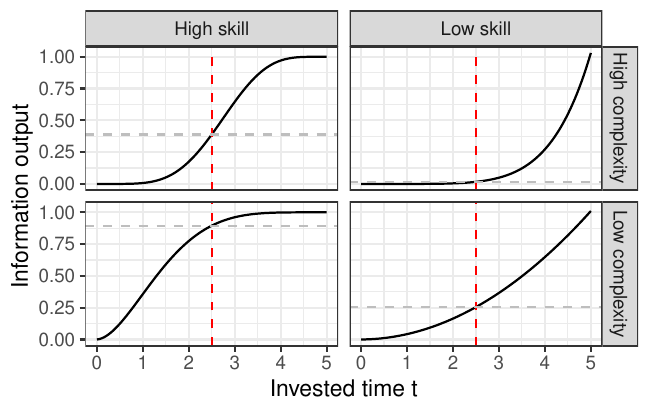}\vspace{-1em}
    \caption{$\text{CDF}_\beta$ for four different configurations}
    \label{fig:cdfbeta}
\end{figure}

\Cref{fig:cdfbeta} visualizes this $\text{CDF}_\beta$ for four separate configurations, contrasting how high and low skill ($s \in \{1; 5\}$) and high and low complexity ($c \in \{2; 6\}$) affect the information $i$ a transfer produces (on the y-axis) given the invested time $t$ (on the x-axis).
When investing $t=2.5h$ (red vertical line) into the transfer, the resulting information output (grey horizontal line) would vastly differ depending on the configuration, ranging from 0.016 (high-complexity transfer performed with low skill, top right) to 0.895 (low-complexity transfer performed with high skill, bottom left).

\begin{figure*}
    \centering
    \includegraphics[width=\linewidth]{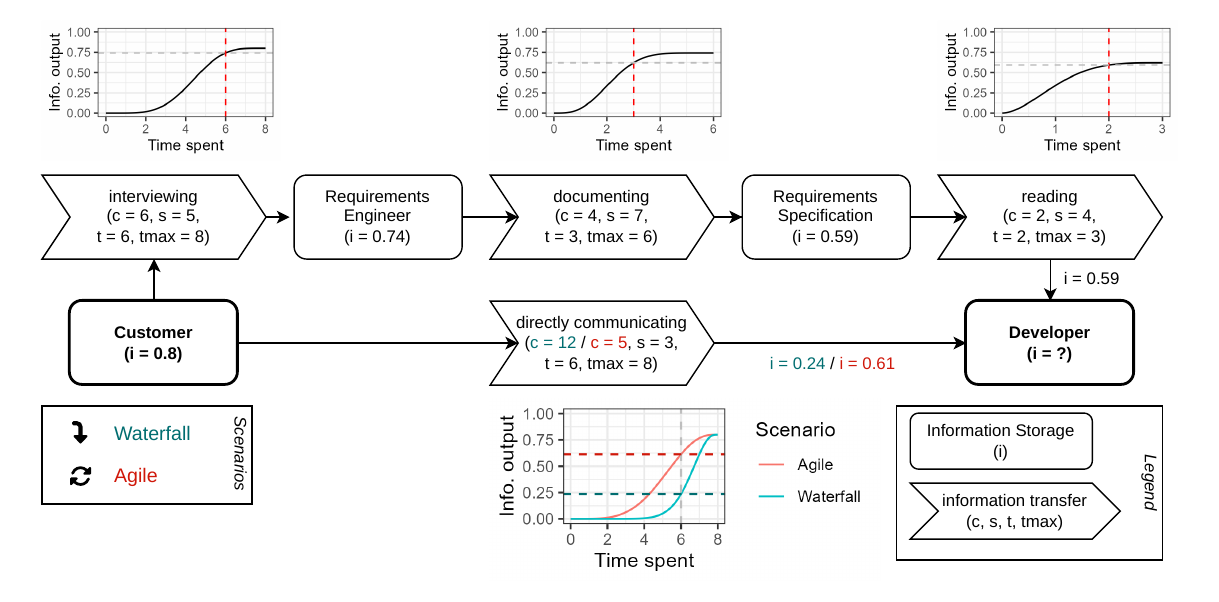}\vspace{-1.5em}
    \caption{Demonstration of Applying the Formalism}
    \label{fig:demonstration}
\end{figure*}

\Cref{fig:demonstration} visualizes the application of the formalization to the scenario described in \Cref{sec:paradigm:anomaly}.
It contains both information storages and transfers with their respective configuration.
Each information transfer is characterized by the $\text{CDF}_\beta$-based function mapping input to output information.
In essence, the figure shows two paths of how the information stored in one customer ($i = 0.8$) can flow to a developer: 

\begin{enumerate}
    \item The \textbf{traditional path}, where information is transferred (1) to the requirements engineer through \textit{interviewing}, (2) then to a specification when \textit{documenting} it, and (3) finally to the developer \textit{reading} this specification.
    \item The \textbf{agile path}, there information is immediately transferred to the developer through \textit{directly communicating}.
\end{enumerate}

The simulation in \Cref{fig:demonstration} configures each transfer activity (interviewing, documenting, reading, and directly communicating) with arbitrary values for complexity $c$, skill $s$, time spent $t$, and maximum time for the activity $t_{max}$.
For example, \textit{interviewing} is considered a more complex activity ($c=6$) than \textit{reading} a specification ($c=2$), and the requirements engineer has a higher skill \textit{interviewing} ($s = 5$) than the developer \textit{directly communicating} with the customer ($s = 3$).
We compare the two flows across two scenarios, a waterfall-like and an agile process.
The difference in the scenarios is operationalized in the complexity of the \textit{directly communicating} task from a $c=12$ (high complexity) to a $c=5$ (medium complexity).
The latter reflects that in an agile scenario, customers are typically more accessible and their feedback is easier to acquire.
This configuration changes the degree of information transferred in the \textit{directly communicating} task, as visualized in the bottom middle graph of \Cref{fig:demonstration} comparing the agile and waterfall information transfer.

Executing the simulation explains the anomaly described in \Cref{sec:paradigm:anomaly}.
In the waterfall-like scenario, where directly communicating with the customer is difficult, the traditional path of transferring information produces more information for the developer than the agile path ($0.59 > 0.24$).
However, when the customer becomes more accessible to the developer (when the complexity of \textit{directly communicating} drops below 6), this situation is reversed ($0.59 < 0.61$).
As a consequence, a developer may bypass the requirements specification despite its high quality, simply because information experiences less resistance on the  alternative path to the target.

This particular resolution determines the flow of information solely based on effectiveness (i.e., on the amount of information arriving at the target).
Determining the flow based on efficiency (i.e., taking cost and resources into account) works similarly.

\section{Discussion}
\label{sec:discussion}

In the following, we discuss opportunities, limitations and challenges of the proposed theory.

\subsection{Opportunities}
\label{sec:discussion:opportunities}

The information-centric paradigm introduced in \Cref{sec:paradigm} provides a foundational perspective that can shift how we empirically study and evaluate quality in RE.
Formalizing the information flow in an organization gives its management the ability to better understand its process and simulate the effect of changes.
This way, an organization can determine the effect of calibrating the RE process.
Connecting this to the example in \Cref{sec:formalization:syntax}, this could mean exploring which information transfer would improve the overall information output given additional time, reducing its complexity, or upskilling the responsible people.

\subsection{Limitations} 
\label{sec:discussion:limitations}

\paragraph{Current limitations and possible extensions of the paradigm}

The paradigm elaborated in \Cref{sec:paradigm} contains but the fundamental concepts necessary to resolve the previously described anomalies. 
However, the information-centric paradigm of RE allows framing several other properties of the RE phase, including:

\begin{itemize}
    \item \textbf{Decay and retention}: Information particles can lose value over time when the system or context changes~\cite{krcmar2015informationsmanagement}. Particularly, intangible information storages may lose information particles more quickly, which may affect the effectiveness of a transfer path. For example, people forget information or participants with intangible domain knowledge are no longer accessible.
    \item \textbf{Value and cost}: The value of a particle may depend on its contribution to the overall body of knowledge, similar to Shannon's information~\cite{shannon1997mathematical}. Modeling particle value by how distinct it is from an existing set of particles and how few information storages contain it could approximate its contribution to the RE process. Furthermore, differentiation of cost, e.g. cost of information acquisition, transformation, delay, or  correction, is necessary.
    \item \textbf{Correctness, errors, misunderstandings}: In our model, the correct, required information is only modeled as present or absent. However, an incorrect or inconsistent information might have a different effect than a missing information. These implications---in particular the risks and consequences of introducing such defects---are not yet included in the presented work.
    \item \textbf{Information merge}: Information storages may receive information particles from multiple sources. For example, a requirements engineer may combine information particles elicited during an interview with particles from their domain knowledge. This may result in  conflicts, which could be modeled in type I formalizations.
    \item \textbf{Prior knowledge}: Considering prior domain knowledge as information particles located in an intangible information storage implies a common information merge with other transfers. We need more research on  understanding this relation in depth as a common instance in RE. 
    \item \textbf{Iterations}: The same information storages may be involved in different information transfers at different points in time. As in agile development, iterative feedback cycles may help transfer more information from customers by presenting them the current solution space~\cite{nuseibeh2001weaving}.    
    \item \textbf{Simulation over time:} We expect that, in future work, type-I simulations would also allow to understand the flow of information over time.
    \item \textbf{Creativity}: Similarly, the RE phase may also unveil previously non-existent requirements by combining several information particles from different sources.
    
\end{itemize}

\paragraph{Limitations of the formalization}
Every type of formalization approach presented in \Cref{sec:formalization:approaches} is inherently limited by the set of assumptions used to frame its syntax.
For example, the presented demonstration in \Cref{sec:formalization:syntax} models information transfers with a specific function (here: $\text{CDF}_\beta$) and a specific set of parameters.
Once again, we emphasize that the current demonstration does not make any claim of representing the real world, it only serves to demonstrate the potential application.

\subsection{Challenges}
\label{sec:discussion:challenges}

Current limitations 
include the difficulty of collecting empirical data that can be used to inductively implement the formalizations.
For a scenario like \Cref{sec:formalization:syntax} to work and a formalization to be useful, its configurations (like the complexity- and skill-values) need to reflect real-world information storages and transfers.
This implies three  steps:

\begin{enumerate}
    \item Identifying the factors that govern information flow in each information transfer step
    \item Collecting data about these factors and their effect on the information flow in type~I formalizations
    \item Abstracting a type~II formalization from these individual cases that models a generic RE process
\end{enumerate}

While requiring considerable effort, we envision that such a path would allow operationalizing the envisioned RE quality paradigm in practice.
Additionally, implementing the information-centric perspective in practice allows to validate the proposed paradigm empirically.

\section{Conclusion}
\label{sec:conclusion}

Requirements are information, and the purpose of RE is to transfer that information from its sources (e.g., stakeholders) to its targets (e.g., developers and testers).
As such, the effectiveness and efficiency of this information flow constitutes RE quality.
Current quality theories of RE have a too narrow, often artifact-centric scope, which produces unexplainable anomalies like developers bypassing high-quality requirements specifications.
Consequently, we propose a paradigm shift in RE quality theories, reframing it from the perspective of information particles and their flow through the RE process. 
We advocate for using this paradigm as a basis for RE quality improvements in research and practice.
We envision that this perspective has the potential to holistically frame RE quality and allow organizations to understand and optimize their RE process in the long-term.

Immediate future work is to empirically track information flow in companies between tangible and intangible information sources. Furthermore, the model requires formalizing of existing processing steps in RE, such as requirements refinement and other aspects discussed in the limitations section.

\noindent\paragraph*{Data Availability Statement}
All our data, scripts, and figures are available online~\cite{renext26replication}.


\bibliographystyle{IEEEtran}
\bibliography{references}

\end{document}